\documentclass[twocolumn]{aastex701}
\usepackage{lipsum}
\usepackage{bm}
\usepackage{amsmath}
\usepackage{amssymb}
\usepackage{graphicx}
\usepackage{subfigure}
\usepackage{hyperref}
\usepackage{color}
\usepackage{comment}
\usepackage{ulem}
\usepackage{CJK}
\usepackage{longtable}
\hypersetup{
	colorlinks=true,
	linkcolor=red,
	citecolor=blue,
}

\allowdisplaybreaks[2]

\usepackage{graphicx}
\usepackage{dcolumn}
\usepackage{bm}

\begin{document}
\nolinenumbers

\title{Black Hole Binary Mergers in Galactic Nuclei: Combined Effects of Secular Dynamics and Stellar Flybys}

\author[0009-0005-7089-1341]{Hao Wang}
\affiliation{Institute for Astronomy, School of Physics, Zhejiang University, 310058 Hangzhou, China}
\email{wanghao\_zju@zju.edu.cn}

\author[0000-0002-0643-8295]{Bin Liu}
\email{liubin23@zju.edu.cn}
\affiliation{Institute for Astronomy, School of Physics, Zhejiang University, 310058 Hangzhou, China}
\affiliation{Center for Cosmology and Computational Astrophysics, Institute for Advanced Study in Physics, Zhejiang University, Hangzhou 310027, China}

\author[0000-0003-0607-8741]{Johan Samsing}
\email{jsamsing@gmail.com}
\affiliation{Center of Gravity, Niels Bohr Institute, Blegdamsvej 17, 2100 Copenhagen, Denmark.}
\affiliation{Niels Bohr International Academy, The Niels Bohr Institute, Blegdamsvej 17, DK-2100, Copenhagen, Denmark}

\begin{abstract}

Galactic nuclei, with the central supermassive black holes (SMBHs) and surrounding dense star clusters, serve as natural laboratories for studying gravitational dynamics in extreme gravitational environments. In this work, we model the long term evolution of black hole binaries (BHBs) in these environments by incorporating the tidal influence of the central SMBH, the potential of the nuclear star cluster, SMBH induced relativistic precession, and stochastic perturbations from stellar flybys. Our numerical simulations show that the long term evolution of BHBs is predominantly governed by the SMBH-driven ZLK mechanism. However, once stellar flybys are included, they can modify both the binary angular momentum and energy. The cumulative effect of repeated flyby encounters substantially reshapes the long term dynamical evolution, eventually leading systems toward either merger or disruption. Based on a population synthesis study, we find that this combined secular and non-secular channel produces a merger rate of approximately $\mathcal{R}\simeq16\mathrm{Gpc^{-3}yr^{-1}}$, which is significantly higher than the rates predicted by dynamical channels that neglect stellar flybys. Furthermore, we predict that a non-negligible fraction of merging systems would retain detectable eccentricities at low frequencies around 5 Hz, thus providing a distinctive observational signature for next generation ground based GW detectors. Our results demonstrate that the stochastic perturbations from the surrounding environment are essential for modeling BHB evolution in galactic nuclei, and that this channel may contribute a major fraction of observed mergers.

\end{abstract}

\keywords{
\uat{Black hole physics}{159} ---
\uat{Gravitational waves}{678} ---
\uat{Galactic nuclei}{604} ---
\uat{Hierarchical triple systems}{1712} ---
\uat{Kozai--Lidov mechanism}{905} ---
\uat{N-body simulations}{1083}
}

\section{Introduction} 
The detection of gravitational waves (GWs) from compact binary coalescences by the LIGO–Virgo–KAGRA (LVK) Collaboration has inaugurated a new era of observational astrophysics (e.g., \citealt{LIGOScientific:2016aoc,KAGRA:2018plz,2026JCAP...03..081A,VIRGO:2014yos}). To date, hundreds of GW events have been reported, the majority of which originate from BHB mergers (e.g.,\citealt{LIGOScientific:2025pvj,2025PhRvD.111f2002C,2025arXiv250818079T}). Despite this rapid progress, the astrophysical origin of merging BHBs remains an open and actively debated problem. While isolated binary stellar evolution can reproduce several observed properties of the BHB population (e.g., \citealt{2012ApJ...759...52D,2016Natur.534..512B,2016MNRAS.460.3545D,Stevenson:2017tfq,Dominik:2013tma,Dominik:2012kk,Dominik:2014yma,Belczynski:2016obo,Belczynski:2010tb,Podsiadlowski:2002ww,PortegiesZwart:1999nm}), growing evidence, including candidates with nonzero eccentricity in the LIGO band and indications of hierarchical mergers, points toward a significant contribution from dynamical formation channels (e.g., \citealt{LIGOScientific:2020iuh,Gayathri:2020coq,Romero-Shaw:2020thy,Gerosa:2021mno,Zevin:2021rtf,Yang:2019cbr,Gerosa:2021mno,Liu:2020gif,Samsing:2020qqd,Stegmann:2025cja,Kiroglu:2025vqy,Paiella:2025qld,Delfavero:2025lup,Bartos:2025pkv,Passenger:2025acb,Gottlieb:2025ugy,OLeary:2005vqo,Miller:2008yw,Banerjee:2009hs,Downing:2009ag,Ziosi:2014sra,Rodriguez:2016kxx,Samsing:2017rat,Samsing:2018isx,Rodriguez:2017pec,Rozner:2022ydm,Liu:2025yok}).

Galactic nuclei provide a particularly promising environment for such dynamical assembly (e.g., \citealt{Baruteau:2010sm,McKernan:2012rf,Mckernan:2017ssq,2017ApJ...835..165B,Stone:2016wzz,Leigh:2017wff,2019ApJ...878...85S,Yang:2019cbr,Liu:2020tcd,Gerosa:2021mno,Ishibashi:2020zzy,Tagawa:2020qll,Tagawa:2020jnc,Tagawa:2021ofj,Li:2021hed,Ford:2021kcw,Samsing:2020tda,Li:2022pnc,Li:2022qoz,Chattopadhyay:2023pil,Su:2025mdd,Rozner:2025iec}). In these regions, BHBs will be affected by a variety of secular effects. A BHB orbiting a central SMBH naturally forms a hierarchical triple system (e.g., \citealt{2012ApJ...757...27A,2017ApJ...846..146P,2019ApJ...875L..31H,2019ApJ...878...58S,2019MNRAS.488...47F}). The outer orbit around the SMBH can induce von Zeipel–Lidov-Kozai (ZLK) oscillations, which periodically drive large-amplitude variations in the inner binary’s eccentricity and inclination (e.g., \citealt{von.Zeipel.1910AN....183..345V,1962P&SS....9..719L,1962AJ.....67..591K,Naoz.2016ARA&A..54..441N}). These oscillations can drive the inner binary to extreme eccentricities, thereby greatly enhancing the efficiency of GW emission. As a result, the inspiral timescale is significantly shortened, enabling mergers that would otherwise exceed a Hubble time. In addition, such eccentricity excitation can leave distinct and potentially observable imprints on the GW signal (e.g., \citealt{Hoang.2018ApJ...856..140H,Lower:2018seu,2021ApJ...921L..31R}).

In the nuclear star cluster, the evolution of ``BHB+SMBH'' systems can
be affected by the additional secular effects. On the one hand, the cluster gravitational potential alters the secular dynamics by inducing extra precession of the inner binary, analogous to the effect of the SMBH, and thereby reshaping the structure of ZLK oscillations (e.g., \citealt{Hamilton.2019MNRAS.488.5489H,Hamilton.2019MNRAS.488.5512H,Hamilton.2021MNRAS.505.4151H}). The cluster potential can drive the system into regimes of extreme eccentricity, substantially enhancing GW emission and facilitating rapid BHB coalescence (e.g., \citealt{Hamilton.2019ApJ...881L..13H}). On the other hand, SMBH introduces additional GR precession effects, including nodal precession and spin-induced precession. These GR effects can drive chaotic orbital dynamics, which broaden the inclination window for BHB mergers, thereby further enhancing the BHB coalescence rate (e.g., \citealt{Liu:2019tqr}).

In dense nuclear environments, binary systems may frequently undergo stellar flybys (e.g., \citealt{Collins.2008AJ....136.2552C,Leigh.2018MNRAS.474.5672L,Michaely.2020MNRAS.498.4924M,Hamilton.2024MNRAS.532.2425H}). These stochastic encounters can induce diffusive changes in orbital parameters, occasionally triggering large eccentricity excursions. Recent study suggest that the combined effect of the smooth cluster potential and discrete stellar encounters can substantially enhance the BHB merger rate compared to models that neglect these environmental influences (e.g., \citealt{Winter-Granic.2024ApJ...973...53W}).

Previous studies have typically focused on limited subsets of the relevant dynamical effects. A comprehensive framework that simultaneously incorporates all secular drivers and non-secular stochastic perturbations over a Hubble time, has yet to be established. In this work, we explore the evolution of BHBs in galactic nuclei by accounting for the combined effects of secular and non-secular dynamical processes, including perturbations from the central SMBH, the smooth potential of the nuclear star cluster, SMBH-induced relativistic precession, and stochastic stellar flybys. Through population-synthesis simulations, we determine the resulting merger rate and predict the eccentricity distribution of merging systems upon entering the sensitivity band of ground-based GW detectors. This approach enables a unified treatment of the complex dynamical environment of galactic nuclei and provides a more complete assessment of their role in producing the observed population of merging black hole binaries.

This paper is organized as follows. In Section \ref{sec:II}, we describe the dynamical framework adopted in this study, including ZLK oscillations, effects of cluster potential, relativistic effects associated with the SMBH, and perturbations from stellar flybys. In Section \ref{sec:III}, we present our numerical methods, initial conditions, evolutionary examples and simulation results, focusing on the long-term evolution and statistical properties of the BHB population. We then estimate the merger rate and discuss observational implications. Finally, we summarize our main conclusions and discuss the implications in Section \ref{sec:IV}.

\section{Dynamical Framework} \label{sec:II}
We model the long-term evolution of BHBs in galactic nuclei by considering a hierarchical three-body system subject to the combined influence of a central SMBH, relativistic effects of SMBH, cluster potential and stochastic stellar flybys. Below we summarize the key dynamical ingredients included in our framework.

\subsection{ZLK Mechanism}
We consider a hierarchical triple black hole system in which two black holes, with masses $m_1$ and $m_2$, form a compact inner binary (BHB), while their center of mass orbits a SMBH of mass $m_3$, thereby constituting the outer binary. The orbital configuration of the hierarchical system is characterized by the semi-major axes of the inner and outer binaries, $a_{\mathrm{in}}$ and $a_{\mathrm{out}}$, their respective eccentricities, $e_{\mathrm{in}}$ and $e_{\mathrm{out}}$, and the mutual inclination $I$ between the inner and outer orbital planes. We further define the total mass $m_{\mathrm{tot}} = m_1 + m_2 + m_3$, and the reduced mass of the inner and outer orbit as $\mu_{\mathrm{in}} \equiv m_1 m_2 / (m_1+m_2)$ and  $\mu_{\mathrm{out}} \equiv (m_1 + m_2)m_3 / m_{\mathrm{tot}}$.

To model the long-term evolution of this system, we adopt double-averaging equations, which average over both the inner and outer orbital periods, and include the Newtonian contribution up to quadrupole order (e.g., \citealt{Liu.2015MNRAS.447..747L,Naoz.2016ARA&A..54..441N}). In this approximation, the evolution of the eccentricity vectors $\boldsymbol{e}_{\mathrm{in}}$, $\boldsymbol{e}_{\mathrm{out}}$ and the dimensionless angular momentum vectors $\boldsymbol{L}_{\mathrm{in}}$, $\boldsymbol{L}_{\mathrm{out}}$ of the inner and outer orbits follows directly from the equations of motion given in Appendix \ref{app:A}, upon setting the cluster mass $M_{\mathrm{cl}}=0$ in Eq. (\ref{eq:A8}) (e.g., \citealt{Hamilton.2019MNRAS.488.5489H,Hamilton.2019MNRAS.488.5512H}). 

While the detailed derivation of the ZLK mechanism is well established in the previous literature and will not be reproduced here, we emphasize the characteristic secular timescale governing these oscillations, given by
\begin{equation}
T_{\rm ZLK}
=
\frac{4}{3}
\frac{a_{\rm out}^{3}}
{\sqrt{G}\,a_{\rm in}^{3/2}}
\frac{\sqrt{m_{1}+m_{2}}}{m_{3}}
\left(1-e_{\rm out}^{2}\right)^{3/2}.
\label{eq:tzlk}
\end{equation}
The BHB also experiences 1PN apsidal precession, which suppresses extreme eccentricity excitation when its characteristic timescale is shorter than the secular timescale. This GR precession timescale is given by (e.g., \citealt{Liu.2015MNRAS.447..747L})
\begin{equation}
T_{\mathrm{GR}}=
\frac{c^2 a_{\mathrm{in}}^{5/2}\left(1-e_{\mathrm{in}}^2\right)}
{3 G^{3/2}(m_1+m_2)^{3/2}}.
\end{equation}
By contrast, GW emission drives the dissipative evolution of the binary eccentricity and angular momentum. The associated merger timescale is (e.g., \citealt{Peters:1963ux,Peters:1964zz})
\begin{equation}
T_{\mathrm{GW}}=\frac{5}{256}\left(\frac{a_{\mathrm{in}}^4 c^5}{G^3 m_1 m_2\left(m_1+m_2\right)}\right)\left(1-e_{\mathrm{in}}\right)^{7 / 2}
\end{equation}
The coupled evolution equations governing the eccentricity and angular momentum under the combined effects of GR precession and GW radiation are provided in Appendix \ref{app:A} (Eqs. (\ref{eq:A11})–(\ref{eq:A14})).

\subsection{Cluster Potential}

In addition to the central SMBH, the BHB is embedded in the extended mass distribution of the surrounding nuclear star cluster. The resulting smooth gravitational potential drives secular evolution of the inner and outer binaries, similar to that induced by the SMBH but with a distinct radial dependence. This axisymmetric potential can substantially modify the long term orbital evolution of the binary system.

Taking the Milky Way as a fiducial example, we model the nuclear star cluster using a Hernquist profile, which provides the axisymmetric potential for our subsequent analysis (e.g., \citealt{Hamilton.2019ApJ...881L..13H,Hamilton.2019MNRAS.488.5489H,Hamilton.2019MNRAS.488.5512H,Hamilton.2021MNRAS.505.4151H}):
\begin{equation}
\Phi_{\mathrm{Hern}}(r)=-\frac{G M_{\text{cl}}}{b_{\mathrm{Hern}}+r},
\end{equation}
where $M_{\text{cl}}$ is the total cluster mass and $b_{\mathrm{Hern}}$ is the scale radius. We adopt $b_{\mathrm{Hern}} = 1\mathrm{pc}$ (e.g., \citealt{Hamilton.2019ApJ...881L..13H}). To obtain the complete gravitational potential of the system, we combine this with the SMBH potential, $\Phi_{\text{SMBH}}$, yielding
\begin{equation}
\Phi = \Phi_{\text{SMBH}} + \Phi_{\text{Hern}}.
\end{equation}

The full set of evolution equations, along with their derivation for a hierarchical triple subject to the combined influence of a SMBH and a stellar cluster, is presented in Appendix~\ref{app:A}. Here we highlight the characteristic secular timescale associated with the joint potential of the SMBH and the cluster, given by 
\begin{equation}\label{eq:6}
T_{\rm Hern+SMBH}
=
\frac{4}{3}
\frac{n_{\rm in}}{G}
\left[
\frac{M_{\rm cl}(3b_{\rm Hern}+r)}
     {r(b_{\rm Hern}+r)^3}
+\frac{m_3}{r^3}
\right]^{-1},
\end{equation}
where $n_{\rm in}=\sqrt{G(m_1+m_2)/a_{\rm in}^3}$ is the mean motion of the inner orbit.

In the limit where the SMBH contribution can be neglected, Eq. (\ref{eq:6}) recovers the cluster-only secular timescale, namely, i.e., $T_{\rm Hern}=T_{\rm Hern+SMBH}$ ($m_3=0$).

\subsection{GR Effects of the SMBH}

When the BHB is sufficiently close to the SMBH, additional GR effects arise that couple the inner orbit, the outer orbit, and the SMBH spin (e.g., \citealt{Liu:2019tqr}). These relativistic effects introduce competing timescales that, together with ZLK oscillations, can drive chaotic orbital evolution. In particular, they modulate the inclination and periapsis non periodically, thereby broadening the inclination window for efficient eccentricity excitation, promoting GW driven mergers. The leading-order GR effects induced by a rotating SMBH includes: 

(i) \textit{Effect I: Lense–Thirring precession of the outer orbit.}
The orbital angular momentum of the outer binary $\boldsymbol{L}_{\mathrm{out}}$, undergoes frame-dragging-induced precession about the SMBH spin vector $\boldsymbol{S}_3$. The spin angular momentum is given by $\boldsymbol{S}_3 = \chi_3 G m_3^2 / c$, where $\chi_3$ is the dimensionless Kerr parameter. For definiteness, we adopt a maximally spinning SMBH $\chi_3 = 1$. The associated precession timescale is
\begin{equation} 
T_{\mathrm{L}_{\mathrm{out}} \mathrm{S}_3}=\frac{2 c^2 a_{\mathrm{out}}^3\left(1-e_{\mathrm{out}}^2\right)^{3 / 2}}{G S_3\left(4+3 (m_1+m_2) / m_3\right)}.
\end{equation}

(ii) \textit{Effect II: de Sitter–like precession of the inner orbit.}
The inner binary angular momentum $\boldsymbol{L}_{\mathrm{in}}$, precesses around $\boldsymbol{L}_{\mathrm{out}}$ due to the relativistic motion of the inner binary in the curved spacetime generated by the SMBH. This effect, analogous to de Sitter (or geodetic) precession, has a characteristic timescale
\begin{equation}
T_{\mathrm{L}_{\mathrm{in}}\mathrm{L}_{\mathrm{out}}}=\frac{2}{3} \frac{c^2 a_{\mathrm{out}}\left(1-e_{\mathrm{out}}^2\right)}{G\left(m_3+\mu_{\mathrm{out}} / 3\right) \left(G M_{\mathrm{tot}} / a_{\mathrm{out}}^3\right)^{1 / 2}}.
\end{equation}

(iii) \textit{Effect III: Lense–Thirring precession of the inner orbit.}
In addition, the inner binary angular momentum vector $\boldsymbol{L}_{\mathrm{in}}$ directly experiences frame-dragging due to the SMBH spin, leading to precession about $\boldsymbol{S}_3$ with timescale
\begin{equation}
T_{\mathrm{L}_{\mathrm{in}} \mathrm{S}_3}=\frac{2 c^2 a_{\mathrm{out}}^3\left(1-e_{\mathrm{out}}^2\right)^{3 / 2}}{G S_3}. 
\end{equation}

The explicit orbit-averaged equations of motion incorporating these GR effects are provided in Appendix \ref{app:A}.

\begin{figure}[htbp!]
\centering
\includegraphics[width=8.4cm,height=12cm]{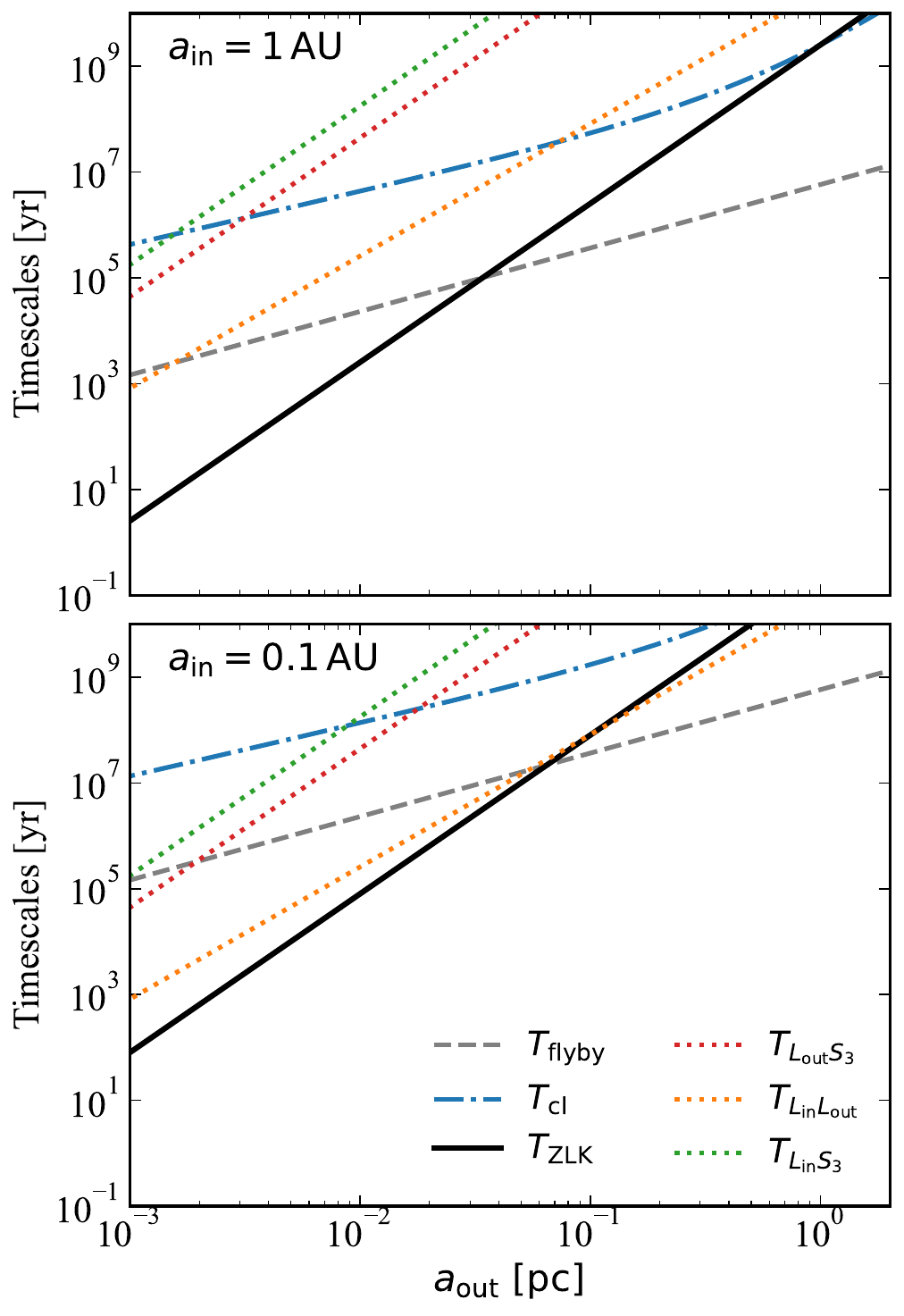}
\caption{\label{FIG:1}Characteristic timescales of different effects (as labeled) as functions of the outer semi-major axis $a_{\mathrm{out}}$. We consider $a_{\mathrm{out}} \in [0.001,2]$ pc with $a_{\mathrm{in}}=1 \text{AU}$ (top panel) and $a_{\mathrm{in}}=0.1 \text{AU}$ (bottom panel). The inner binary consists of equal-mass components, $m_1=m_2=15 M_{\odot}$, and the tertiary has mass $m_3=4\times10^{6}M_{\odot}$. A low eccentricity outer orbit is assumed ($e_{\mathrm{out}}=0.001$).
}
\end {figure}

\subsection{Stellar Flybys}

In a dense stellar environment, a single flyby can reset a binary's orbit, alter its orientation, and drive it toward extreme eccentricities (e.g., \citealt{1975AJ.....80..809H,1975AJ.....80.1075H,1975MNRAS.173..729H}). Over multiple encounters, these perturbations cause a random walk in the binary's angular momentum, which may ultimately lead to merger or disruption (e.g., \citealt{Winter-Granic.2024ApJ...973...53W}). 

The characteristic timescale between successive flyby events is given by
\begin{equation}\label{eq:10}
T_{\mathrm{flyby}}=\frac{1}{n_{\star} \sigma_p v_p},
\end{equation}
where $\sigma_p = \pi b_{\max}^2$ is the geometric cross section for encounters, and $n_{\star}$ denotes the stellar number density. We adopt $b_{\max} = 5 a_{\mathrm{in}}$ and
\begin{equation}
n_{\star}(r)=\frac{2.8 \times 10^6 M_{\odot} \mathrm{pc}^{-3}}{\left\langle m_{\star}\right\rangle} \left(\frac{r}{0.22 \mathrm{pc}}\right)^{-\gamma},
\end{equation}
with $\gamma = 1.2$ for $r < 0.22$ pc and $\gamma = 1.75$ for $r > 0.22$ pc. The mean stellar mass is taken to be $\left\langle m_{\star} \right\rangle \approx 1 M_{\odot}$.

The characteristic flyby velocity is set by the velocity dispersion in the galactic nucleus, given by
\begin{equation}\label{eq:12}
\begin{aligned}
& \sigma(r)= \begin{cases}280 \mathrm{~km} \mathrm{~s}^{-1} \sqrt{0.1 \mathrm{pc} / r} \sqrt{1-0.035(r / 0.1 \mathrm{pc})^{2.2}}, \\
250 \mathrm{~km} \mathrm{~s}^{-1} \sqrt{0.1 \mathrm{pc} / r}. \end{cases}
\end{aligned}
\end{equation}

Fig. \ref{FIG:1} compares the timescales associated with each dynamical ingredient in our model for a BHB and SMBH system embedded in a nuclear star cluster. We see that all characteristic timescales exhibit an approximately linear dependence on the outer semi-major axis $a_{\mathrm{out}}$, across the parameter range considered. 

In the top panel, the ZLK mechanism dominates the secular evolution at small galactocentric distances, where the tidal perturbation from the SMBH is strongest. As $a_{\mathrm{out}}$ increases, the ZLK timescale lengthens, and stellar flybys gradually take over as the primary dynamical driver due to their shorter characteristic timescale and cumulative impact. Throughout this range, the smooth cluster potential and most SMBH induced relativistic effects remain subdominant. When $a_{\mathrm{in}}=0.1 \text{AU}$, as shown in the bottom panel, the ZLK mechanism again governs the evolution at small $a_{\mathrm{out}}$, while flybys become progressively more important at larger separations. One notable difference arises in the near field regime, where the SMBH induced de Sitter precession operates on a timescale shorter than that of stellar flybys, thereby exerting a stronger influence on the BHB orbital evolution.

To incorporate stellar flybys into our secular evolution framework, we adopt an impulsive approximation. Because the outer orbital angular momentum is much larger than that of the inner binary ($L_{\mathrm{out}} \gg L_{\mathrm{in}}$), individual encounters primarily perturb the inner orbit while leaving the outer orbit largely intact. We implement this approach using the \texttt{REBOUND} N-body package (e.g., \citealt{Rein.2012A&A...537A.128R}). At each flyby event, we calculate the instantaneous change in the inner binary's eccentricity and angular momentum, and then update these quantities accordingly \setcounter{footnote}{0}\footnote{In the flyby simulations, the perturbing black hole is initialized at a sufficiently large distance from the target binary so that its initial gravitational influence is negligible. The corresponding initial velocity is calculated from Eq. (\ref{eq:12}) }. Between successive flybys, we evolve the hierarchical triple system using the secular equations provided in Eq. (\ref{eq:A8}) of Appendix \ref{App:A}.

\section{Numerical Simulations} \label{sec:III}

\subsection{System setup}

For the initial conditions of the numerical simulations, we adopt equal component masses of $15M_{\odot}$ for the inner BHB, while the central SMBH is assigned a mass of $4\times10^{6}M_{\odot}$. The total mass of the surrounding stellar cluster is set to $3.2\times10^{7}M_{\odot}$, corresponding to eight times the SMBH mass, consistent with typical nuclear star cluster environments.

The initial semi-major axes of the inner and outer orbits are sampled from log-uniform distributions, with $[0.1, 50]\mathrm{AU}$ and $[0.001, 2]\mathrm{pc}$. The initial eccentricities of both the inner and outer orbits are drawn independently from the thermal distribution $p(e)=2 e(0 \leq e<1)$. The orbital inclination $I$ is drawn from an isotropic distribution, uniform in $\cos I$ over $[-1,1]$. The arguments of periapsis, $\omega_{\mathrm{in}}$ and $\omega_{\mathrm{out}}$, and the longitudes of ascending nodes $\Omega_{\mathrm{in}}$, are independently sampled from uniform distributions over $[0, 2\pi]$. The misalignment angle between the SMBH spin and the outer orbital angular momentum is fixed at $30^\circ$ to incorporate the leading-order relativistic spin-orbit coupling effects of the SMBH.

The stability condition is given by (e.g., \citealt{Mardling.2001MNRAS.321..398M})
\begin{equation}
\begin{aligned}
\frac{a_{\mathrm{out}}}{a_{\mathrm{in}}(1+e_{\mathrm{in}})} >& 2.8\left(1+\frac{m_3}{(m_1+m_2)}\right)^{2/5}
\\ &\times\frac{(1+e_{\mathrm{out}})^{2/5}}{(1-e_{\mathrm{out}})^{6/5}}
 \left(1-\frac{0.3 I}{180^\circ}\right).   
\end{aligned}
\end{equation}

Based on these distributions, we generate $10^{6}$ realizations of initial conditions and subsequently apply the hierarchical three-body stability criterion, retaining $53665$ dynamically stable systems.

The numerical integration of Eq.~(\ref{eq:A8}) is carried out up to Hubble Time ($10\,\mathrm{Gyr}$), or until one of the following termination criteria is satisfied:

(i) \textit{Disruption}: After each stellar flyby, the stability criterion is re-evaluated. If violated, the system is classified as dynamically disrupted.

(ii) \textit{Merger}: The inner binary is assumed to merge rapidly once its GW inspiral timescale becomes shorter than the outer orbital period, $T_{\mathrm{GW}} < P_{\mathrm{out}}$, where $P_{\mathrm{out}} = 2\pi \sqrt{a_{\mathrm{out}}^{3}/(G m_{\mathrm{tot}})}$. Another condition for merger is its semi-major axis decreases to $1\%$ of its initial value.
To illustrate the condition for the former, consider a representative configuration with $a_{\mathrm{in}}=10 \mathrm{AU}$ and $a_{\text {out }}=0.1 \mathrm{pc}$. In this case, the merger criterion imposes a requirement on the inner BHB eccentricity $1-e_{\mathrm{in}} \lesssim 10^{-4}$.

\subsection{Examples of orbital evolution}

We present two representative evolution examples illustrating the time evolution of orbital elements eccentricity $e_{\text {in }}$, semi-major axis $a_{\text {in }}$, and inclination $I$ for BHBs at two different galactocentric distances $a_{\text {out }}$ which can be classified into near field and far field.

\begin{figure*}[htbp!]
\centering
\includegraphics[width=18cm,height=18cm]{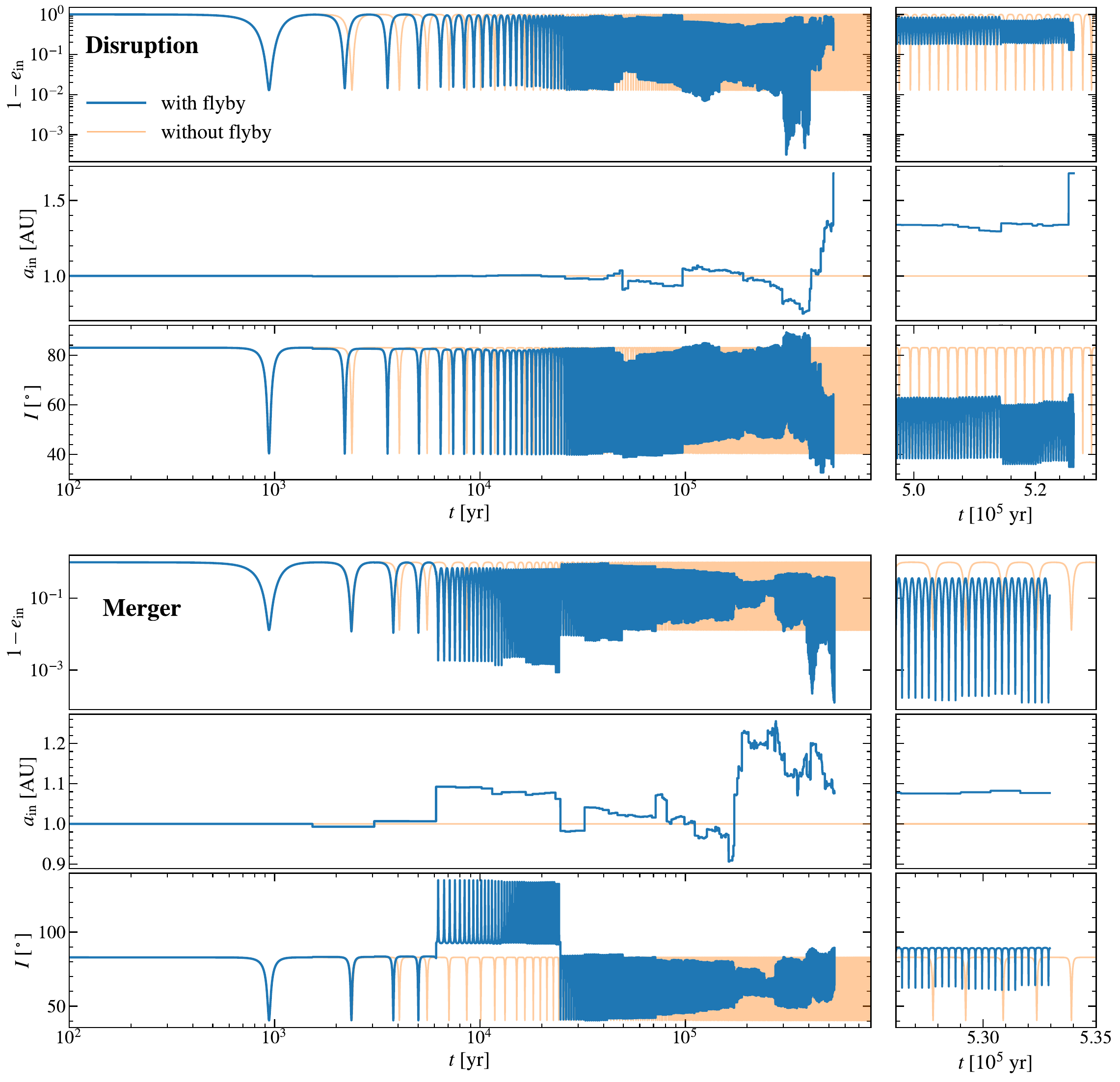}
\caption{\label{FIG:2}
Evolution of the inner binary orbital elements eccentricity $e_{\mathrm{in}}$, semi-major axis $a_{\mathrm{in}}$, and inclination $I$ for a near-field BHB system subject to the combined influence of a central SMBH, the surrounding stellar cluster, relativistic precession, and stochastic stellar flybys. The upper three panels shows a disruption outcome, while the lower three panels illustrates a merger event.
Both panels adopt identical initial conditions. The inner binary consists of equal-mass black holes with $m_1 = m_2 = 15\,M_\odot$, orbiting an SMBH of mass $m_3 = 4 \times 10^{6}\,M_\odot$, embedded in a stellar cluster of mass $M_{\mathrm{cl}} = 3.2 \times 10^{7}\,M_\odot$, modeled with a Hernquist profile of scale radius $b_{\mathrm{Hern}} = 1\,\mathrm{pc}$.
The initial orbital configuration is specified by $\omega_\text{in} = 90^\circ$, $\omega_\text{out} = 0^\circ$, and $\Omega_\text{in} = 0^\circ$. The SMBH spin axis is inclined by $30^\circ$ with respect to the outer orbital angular momentum. 
The outer orbit is initialized with $a_{\mathrm{out}} = 10^3\,\mathrm{AU}$ and $e_{\mathrm{out}} = 10^{-3}$, while the inner binary has $a_{\mathrm{in}} = 1\,\mathrm{AU}$ and $e_{\mathrm{in}} = 10^{-3}$. The mutual inclination between the inner and outer orbits is $I = 83^\circ$, placing the system in the regime of efficient secular excitation. For comparison, we also show the corresponding orbital evolution in the absence of stellar flybys. The right hand panels present zoom-in views of the final stages of the evolution.
}
\end {figure*}

Fig. \ref{FIG:2} shows the near-field orbital evolution leading to disruption (upper three panels) and merger (lower three panels) for a BHB with $a_{\mathrm{out}}=1000\,\mathrm{AU}=0.00485\,\mathrm{pc}$. 
For comparison, we show the corresponding orbital evolution in the absence of stellar flybys. 

\begin{figure*}[htbp!]
\centering
\includegraphics[width=18cm,height=18cm]{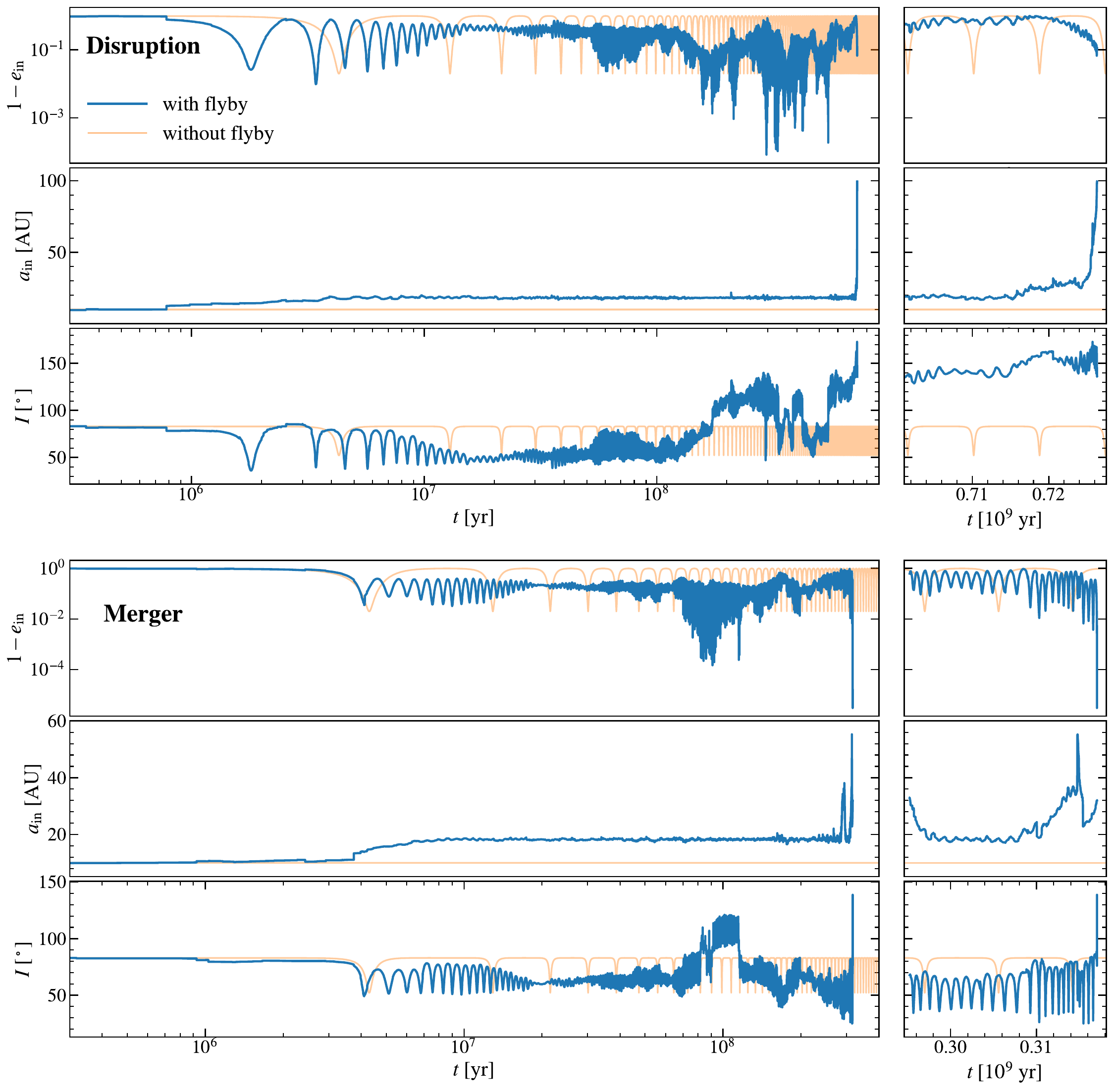}
\caption{\label{FIG:3}
Same as Fig.~\ref{FIG:2}, but for a wider outer orbit with $a_{\mathrm{out}} = 0.3\,\mathrm{pc}$ and $a_{\mathrm{in}} = 10\,\mathrm{AU}$. 
}
\end {figure*}

The characteristic timescales of the ZLK mechanism and stellar flybys are $\sim 2.9\times 10^2\,\mathrm{yr}$ and $\sim 9.7\times10^3\,\mathrm{yr}$. The ZLK oscillation therefore dominates the secular evolution, while stellar flybys provide stochastic perturbations superimposed on the secular dynamics. Other dynamical effects operate on significantly longer timescales (see Fig. \ref{FIG:1}). In particular, only Effect II, with a characteristic timescale of $\sim 4.2\times10^4\,\mathrm{yr}$, becomes partially comparable and contributes non-negligibly, whereas all remaining processes are dynamically subdominant.
As shown in Fig.~\ref{FIG:2}, the ZLK mechanism induces periodic oscillations in eccentricity, while stochastic perturbations from stellar flybys intermittently drive the system to extremely high eccentricities, resulting in irregular and quasi-random orbital behavior. The semi-major axis $a_{\mathrm{in}}$ and inclination $I$ also exhibit stochastic variations, with $a_{\mathrm{in}}$ showing a secular trend of gradual growth.

The top and bottom panels correspond to 345 and 428 stellar flyby events, respectively.  In the top three panels of Fig.~\ref{FIG:2}, the gradual increase in $a_{\mathrm{in}}$ ultimately leads to the disruption of the hierarchical configuration. In contrast, in the bottom three panels, the system undergoes orbital flips and maintains a nearly constant $a_{\mathrm{in}}$, while the eccentricity is driven beyond $1-e_{\mathrm{in}} \lesssim 10^{-4}$, satisfying the merger criterion. The disruption and merger occur at $\sim 5.2\times10^5\,\mathrm{yr}$ and $\sim 5.3\times10^5\,\mathrm{yr}$, respectively.
In the absence of stellar flybys, the system exhibits regular secular evolution characterized by quasi-periodic oscillations in both eccentricity and mutual inclination, while the semi-major axis remains nearly unchanged. This behavior is consistent with the expected signature of ZLK-driven dynamics.  

Fig.~\ref{FIG:3} presents the corresponding evolution for a far-field system with $a_{\mathrm{out}} = 0.3\,\mathrm{pc}$. The characteristic timescales of the ZLK mechanism and stellar flybys are $\sim 2.2 \times10^6\,\mathrm{yr}$ and $\sim 1.4\times10^4\,\mathrm{yr}$, respectively. In contrast to the near-field case shown in Fig.~\ref{FIG:2}, the ZLK timescale is now more than two orders of magnitude longer than the flyby timescale, indicating that stochastic stellar encounters dominate the orbital evolution. While the qualitative dynamical behavior remains similar, the system is significantly more stable against perturbations from stellar flybys, allowing for longer survival timescales. In both cases for disruption and merger, the systems experience more than $10^5$ flyby encounters before either disruption or merger occurs. The disruption and merger occur at $\sim 7.3\times10^8\,\mathrm{yr}$ and $\sim 3.2\times10^8\,\mathrm{yr}$, respectively. In this regime, GR precession operates on timescales that are sufficiently long to be dynamically negligible. The BHB therefore evolves under the combined influence of the potential of SMBH and tellar cluster, and repeated stellar flybys.

\subsection{Results}

Among the 53665 numerical realizations that satisfy the stability criterion, we find that 24.8\% of systems merge, 55.6\% are disrupted, and the remaining 19.6\% remain stable over a Hubble time. These statistics indicate that BHBs embedded in hierarchical triples within galactic nuclei rarely remain dynamically stable over cosmological timescales.

\begin{figure*}[htbp!]
\centering
\includegraphics[width=18cm,height=11cm]{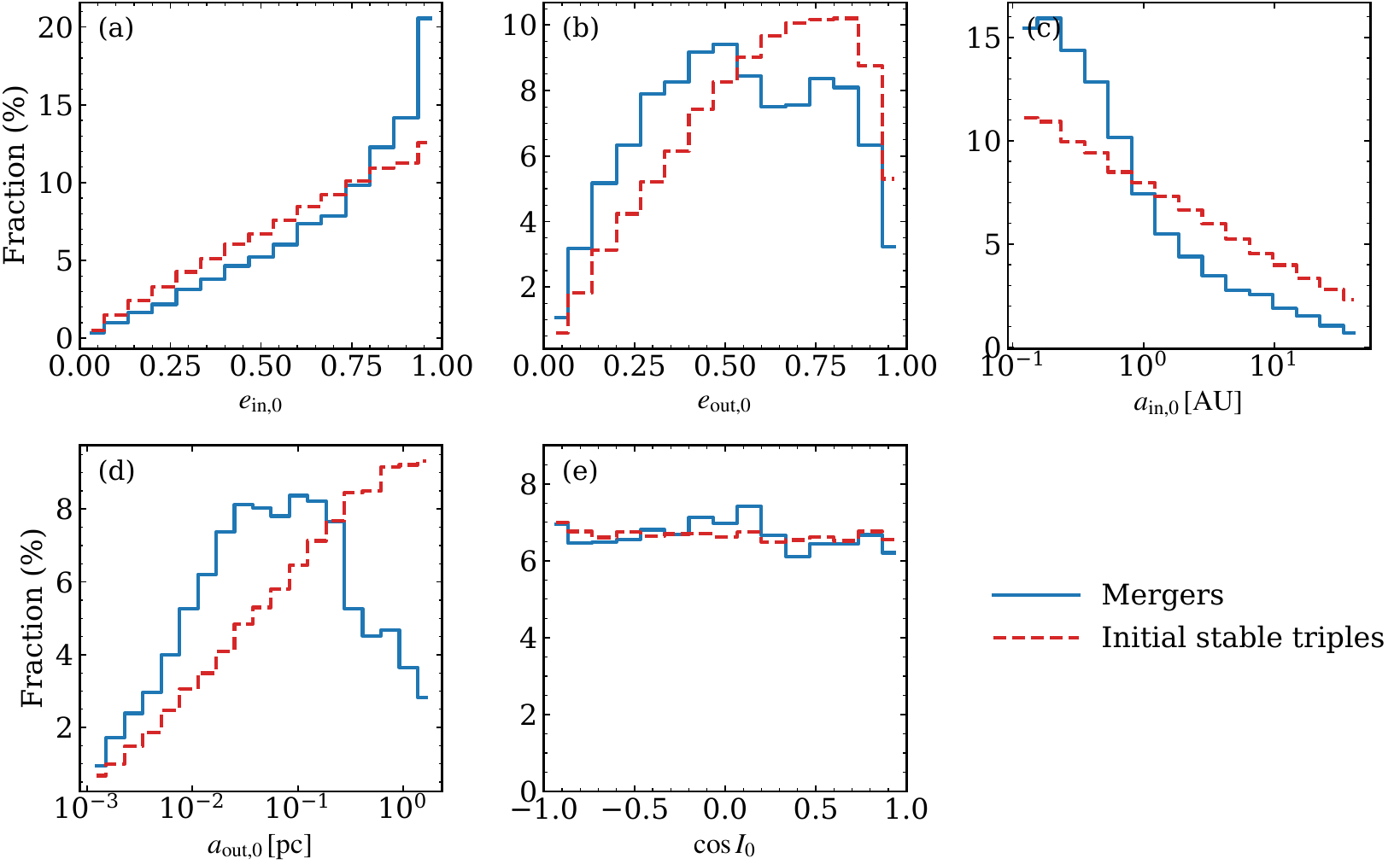}
\caption{\label{FIG:4}Parameter distributions of SMBH+BHB triples studied in our population-synthesis simulations. The blue solid line shows the fraction of merging BHBs. The red dashed line indicates the initial stable triples obtained from simulations that have been filtered by the stability condition.}
\end {figure*}

Building on this ensemble, we extract the distributions of key orbital parameters for the merging BHB population, including the semi-major axes, eccentricities, and mutual inclinations. These distributions quantify how the merger probability depends on the initial orbital configuration and identify the regions of parameter space that are most favorable for BHB coalescence. The detailed results are presented in Fig.~\ref{FIG:4}. For comparison, we also show the corresponding distributions of the dynamically stable initial population after applying the hierarchical stability criterion.

In panel (a), systems with larger initial inner eccentricities exhibit a higher merger fraction compared with the stable initial population. This trend arises because highly eccentric binaries have shorter GW inspiral timescales and can more readily satisfy the merger criterion once their eccentricity is further enhanced by dynamical perturbations. The transition occurs at approximately $e_{\mathrm{in},0}\sim0.75$; below this value, the merger distribution closely follows the underlying stable initial population.

Panel (b) shows that merging systems are preferentially drawn from binaries with lower outer eccentricities compared with the stable population. Highly eccentric outer orbits lead to stronger variations in the perturbing tidal field and shorter relativistic and secular precession timescales, which can accelerate the disruption of the hierarchical configuration and reduce the probability of merger before the system is destabilized.

In panel (c), binaries with smaller initial inner semi-major axes exhibit an enhanced merger fraction. This is expected because tighter binaries require less eccentricity excitation to enter the efficient gravitational-wave-driven inspiral regime and can therefore merge more readily following environmental perturbations. The transition occurs at approximately $a_{\mathrm{in},0}\sim1\mathrm{AU}$.

Panel (d) demonstrates that the merger fraction is enhanced at small and intermediate outer semi-major axes but decreases in the far-field region. At large $a_{\mathrm{out}}$, the characteristic dynamical timescales become longer, reducing the efficiency of both secular excitation and stochastic perturbations in driving binaries toward merger within a Hubble time.

Finally, panel (e) reveals a pronounced enhancement of the merger fraction near a mutual inclination of $I_{0}\sim90^\circ$. This behavior reflects the contribution of the ZLK mechanism, for which nearly perpendicular orbital configurations maximize angular momentum exchange between the inner and outer orbits and facilitate extreme eccentricity excitation.

A key finding of our numerical simulations is that stellar flyby encounters play a dominant role in both merger and disruption pathways. This conclusion is supported by the fact that $\sim90\%$ and $\sim77\%$ of merger and disruption events are triggered immediately following a flyby encounter, as identified by the adopted termination criteria. In contrast, only $\sim10\%$ and $\sim23\%$ of systems reach the merger and disruption thresholds through purely secular evolution occurring between successive flybys. These results highlight the critical role of stochastic stellar perturbations in driving the long-term dynamical evolution of BHBs in dense nuclear environments.

\begin{figure}[htbp!]
\centering
\includegraphics[width=8cm,height=8cm]{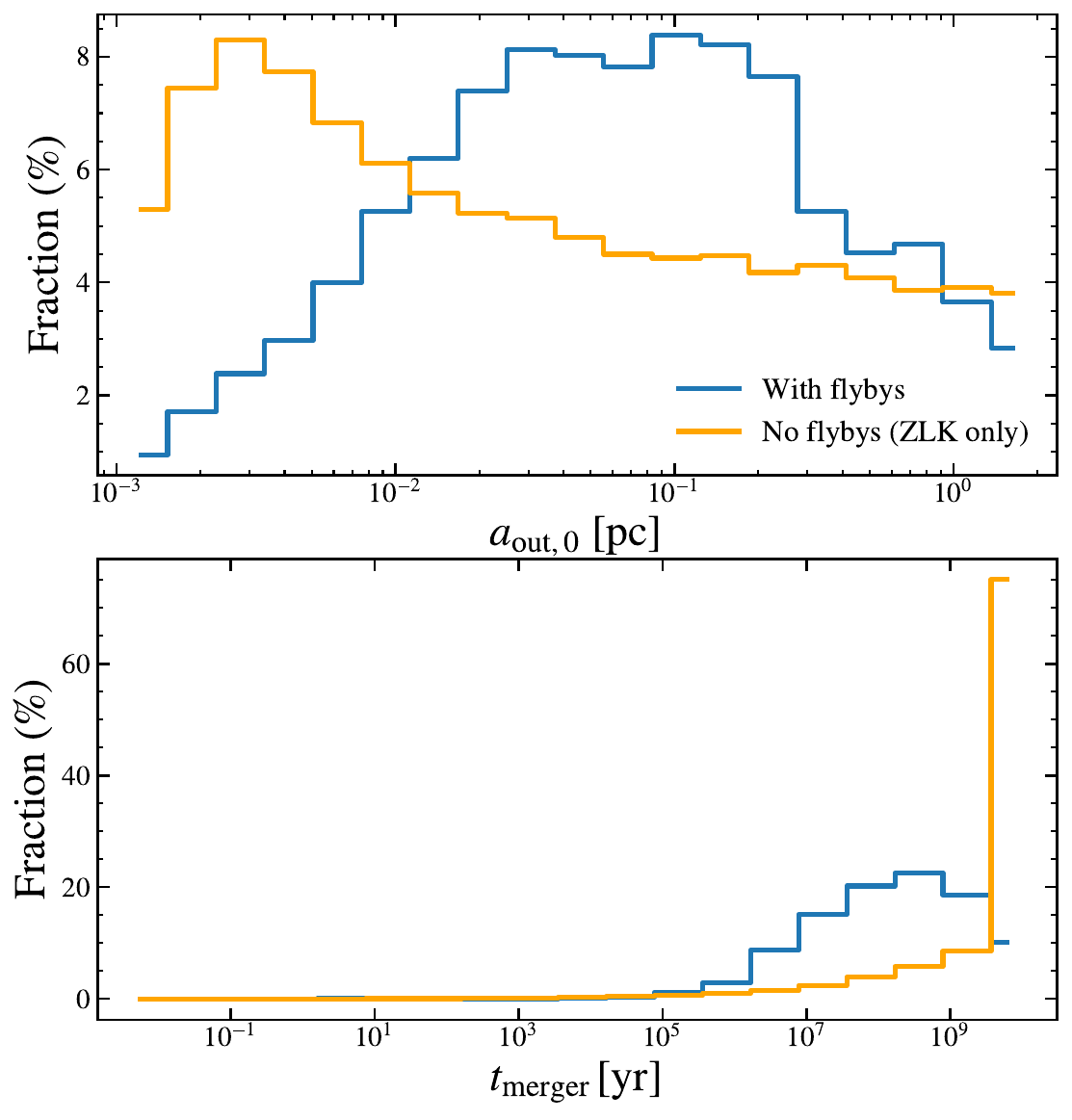}
\caption{\label{FIG:5}
Upper panel: BHB merger fraction as a function of the outer semi-major axis, $a_{\mathrm{out}}$.
Lower panel: Distribution of merger time for BHB systems that successfully merge. 
In both panels, the results from purely ZLK-driven evolution are shown for comparison.
All curves are normalized and therefore represent relative distributions rather than the absolute number of merger events.}
\end{figure}

To isolate the impact of non-secular perturbations, particularly stochastic stellar flybys, we compare our results with a reference population in which BHB evolution is driven solely by secular dynamics, dominated by the ZLK mechanism. The upper panel in Fig.~\ref{FIG:5} illustrates the merger fraction of BHBs as a function of outer semi-major axis, $a_{\mathrm{out}}$. The blue solid line corresponds to the results shown in panel (d) of Fig.~\ref{FIG:4}.
 For comparison, we also show the merger fraction expected under purely ZLK-driven evolution (see the orange line). 

To quantify the merger fraction expected from purely secular ZLK evolution, we generated $10^6$ realizations of the initial parameter distributions described above. After applying the hierarchical stability criterion, 535617 dynamically stable systems were retained for further analysis.
For each stable realization, we estimated the merger time using the analytical prescription of \citet{Liu.2018ApJ...863...68L}, which describes BHB evolution driven exclusively by secular ZLK oscillations. In the limit where octupole-order effects are negligible, the merger timescale can be approximated as
\begin{equation}
T_{\mathrm{m}} \simeq T_{\mathrm{m}, 0}\left(1-e_{\mathrm{max}}^2\right)^3,
\end{equation}
where $T_{\mathrm{m},0}=T_{\rm GW}(e_{\rm in}=0)$ is the circular GW inspiral timescale, and $e_{\max}$ is the maximum eccentricity attained during ZLK-driven evolution.

The maximum eccentricity can be obtained from the conservation of total angular momentum and secular energy. Including the leading-order effect of relativistic apsidal precession, $e_{\max}$ is determined implicitly through
\begin{equation}
\begin{aligned}
& \frac{3}{8} \frac{j_{\min }^2-1}{j_{\min }^2}\left[5\left(\cos I_0+\frac{\eta}{2}\right)^2\right. \\
& \left.\quad-\left(3+4 \eta \cos I_0+\frac{9}{4} \eta^2\right) j_{\min }^2+\eta^2 j_{\min }^4\right] \\
& \quad+\varepsilon_{\mathrm{GR}}\left(1-j_{\min }^{-1}\right)=0
\end{aligned}
\end{equation}
where
\begin{equation}
j_{\min}
\equiv
\sqrt{1-e_{\max}^{2}},
\end{equation}
and
\begin{equation}
\eta=\frac{\mu_{\mathrm{in}}}{\mu_{\mathrm{out}}}\left[\frac{(m_1+m_2) a}{m_{\text {tot }} a_{\mathrm{out}}\left(1-e_{\mathrm{out}}^2\right)}\right]^{1 / 2} ,
\end{equation}
is the ratio of the inner to outer orbital angular momentum evaluated for a circular inner binary. The dimensionless parameter
\begin{equation}
\varepsilon_{\mathrm{GR}}=\frac{3 G (m_1+m_2)^2 (a_{\mathrm{out}} \sqrt{1-e_{\mathrm{out}}^2})^3}{c^2 a_{\mathrm{in}}^4 m_3}
\end{equation}
quantifies the strength of general relativistic pericenter precession relative to the quadrupole-order ZLK torque. The merger fraction in the pure ZLK scenario is then obtained by identifying systems with $T_{\mathrm{m}} < 10\mathrm{Gyr}$.

Among all dynamically stable systems, the analytical calculation gives a merger fraction of approximately \(7.6\%\) for the pure-ZLK case. In comparison, our fiducial model yields a merger fraction of \(24.8\%\), corresponding to an enhancement by a factor of \(\sim3\) relative to the flyby-free case.

The upper panel of Fig. \ref{FIG:5} reveals a clear trend: BHBs located at intermediate distances about $0.02-0.1 \mathrm{pc}$ from the SMBH exhibit significantly higher merger fractions when stellar flybys are included. This enhancement reflects the cumulative impact of stochastic perturbations in the intermediate regions of galactic nuclei. 
In contrast, the merger fractions in the inner ($< 0.02 \mathrm{pc}$) and outer ($>0.1\mathrm{pc}$) regions are relatively small. In the outer region, the reduced merger efficiency is primarily caused by the longer characteristic dynamical timescales, which limit the ability of perturbations to drive mergers within a Hubble time. In the inner region, the lower merger fraction mainly results from the adopted initial parameter distribution, which contains fewer systems capable of reaching the merger regime. These combined effects naturally produce a peak in the merger fraction at intermediate distances. 

The pure ZLK scenario, however, yields a markedly different spatial dependence, with relatively low merger fractions at intermediate and large distances, and higher merger efficiency only in the inner region. The peak of the merger fraction is around $0.003 \mathrm{pc}$. The merger fractions are normalized independently at each separation bin. However, when adopting a common normalization corresponding to the same number of initial systems, the purely ZLK-driven channel produces substantially fewer mergers than the flyby-assisted evolution. This demonstrates that stellar flybys can significantly enhance the overall BHB merger efficiency. 

Lower panel of Fig.~\ref{FIG:5}  presents the distribution of merger time for BHBs that successfully coalesce. 
The comparison demonstrates that stellar flybys substantially shorten BHB merger times in galactic nuclei. For the mergers induced by the pure ZLK effect, the majority of systems fail to merge within a Hubble time, with only $\sim 7.6\%$ reaching coalescence within the age of the Universe, predominantly in the range $10^9$--$10^{10}\,\mathrm{yr}$. By contrast, when flyby perturbations are included, stochastic gravitational interactions efficiently excite the eccentricity and accelerate the dissipation of orbital energy and angular momentum, thereby significantly enhancing the merger efficiency. The peak of its merger time distribution is around $10^8 \mathrm{yr}$. 

\subsection{Residual eccentricity}

For eccentric BHBs, the GW emission is not monochromatic but instead spread over a spectrum of harmonics of the orbital frequency, with the emission concentrated near pericenter passages. The characteristic peak frequency associated with each pericenter passage is given by (e.g., \citealt{Wen.2003ApJ...598..419W})
\begin{equation}
f_{\mathrm{GW}}^{\text{peak}}=\frac{\sqrt{G\left(m_1+m_2\right)}}{\pi}\frac{(1+e_{\mathrm{in}})^{1.1954}}{\left[a_{\mathrm{in}}\left(1-e_{\mathrm{in}}^2\right)\right]^{3/2}}.
\end{equation}

In our numerical simulations, all merger events are triggered by eccentricity excitations. Although GW radiation tends to circularize the binary orbit during the late inspiral, the orbital eccentricity may not be completely damped by the time the GW frequency enters the sensitivity band of ground based detectors.  

\begin{figure}[htbp!]
\centering
\includegraphics[width=8cm,height=8cm]{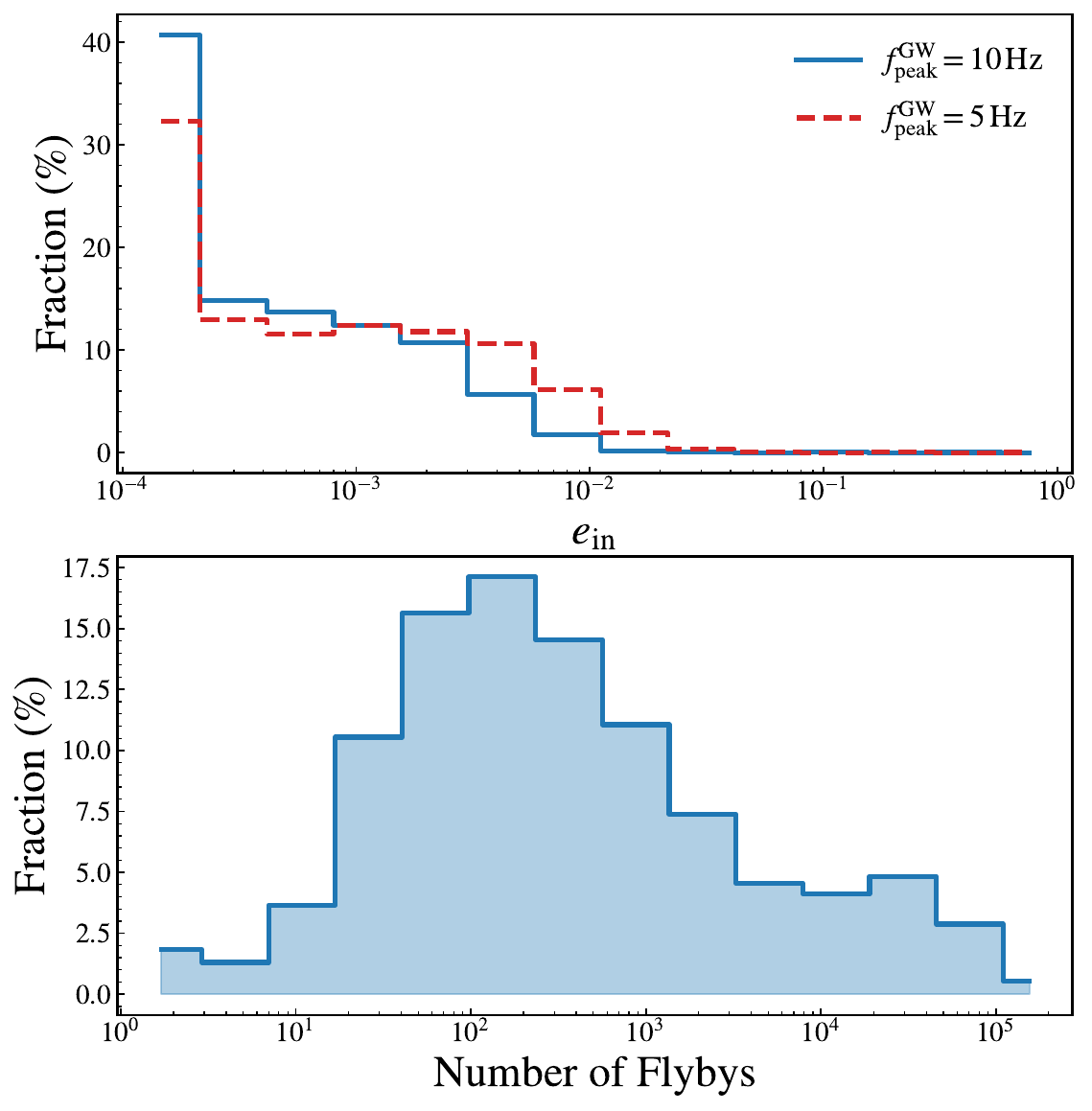}
\caption{\label{FIG:6}
Upper panel: Distribution of the residual eccentricities of merging BHBs when entering the GW frequency band, evaluated at 10 Hz and 5 Hz thresholds, respectively. Lower panel: Distribution of the number of stellar flyby encounters experienced by each merging BHB throughout its evolutionary history.
}
\end{figure}

Fig. \ref{FIG:6} presents the resulting residual eccentricity distributions evaluated at two reference frequencies, 10 Hz and 5 Hz, corresponding to the sensitive bands of current and next generation ground based GW detectors, respectively \footnote{Using the orbital parameters $a_{\mathrm{in}}$ and $e_{\mathrm{in}}$  at the onset of merger, we further infer the eccentricity of BHBs in isolation as they enter the detector band, employing the Peters formalism (e.g., \citealt{Peters:1963ux,Peters:1964zz}).}.
The former corresponds to the typical low-frequency cutoff of current detectors such as LIGO, while the latter approximates the extended sensitivity expected for next-generation facilities such as the Einstein Telescope (e.g., \citealt{2026JCAP...03..081A}) and Cosmic Explorer (e.g.,\citealt{Reitze:2019iox}).

For the two scenarios shown in the figure, the fraction of BHBs gradually decreases as the eccentricity increases.
For the $10\,\mathrm{Hz}$ threshold, we find that $\sim 77\%$ of BHBs have already undergone significant orbital circularization upon entering the detector band, with eccentricities effectively indistinguishable from zero ($e_{\mathrm{in}}\lesssim10^{-3}$). Approximately $23\%$ retain mild eccentricities in the range $10^{-3}\lesssim e_{\mathrm{in}} \lesssim 10^{-2}$, while systems with $e_{\mathrm{in}} \gtrsim 10^{-2}$ are exceedingly rare, with only 0.3\% realizations in our sample. 
At the lower threshold of $5\,\mathrm{Hz}$, because the semi-major axis and eccentricity are larger compared to the 10Hz case, the fraction of systems retaining measurable eccentricity increases modestly. In this case, $\sim 33\%$ of BHBs exhibit eccentricities in the range $10^{-3}\lesssim e_{\mathrm{in}}\lesssim 10^{-2}$, while $\sim 2\%$ reach $e_{\mathrm{in}}\gtrsim 10^{-2}$. These results suggest that next-generation ground-based detectors, with improved low-frequency sensitivity, will have an enhanced capability to detect eccentric BHB mergers originating from hierarchical triple systems in galactic nuclei.

Lower panel of Fig.~\ref{FIG:6} shows the distribution of the number of stellar flyby events experienced by merging BHBs. More than half of these systems undergo in excess of $10^2$ flybys, indicating sustained perturbative interactions throughout their evolution. The peak of the flyby number distribution is around 137, and the distribution shape is large in the middle and small at both ends. In extreme cases, the cumulative number of flybys reaches $\sim 10^{5}$, underscoring the importance of repeated weak encounters in driving the system toward merger.

\subsection{Merger Rate} 
We now turn to the BHB merger rate, defined as the number of merging events per unit volume per unit time ($ \mathrm{pc}^{-3} \mathrm{yr}^{-1}$) in the local universe. This rate can be estimated by (e.g., \citealt{Hamilton.2019ApJ...881L..13H}),
\begin{equation}\label{eq:20}
\mathcal{R}=Y_{\text{form}}\, M_{\mathrm{cl}}\, n_{\mathrm{cl}}\, f_{\mathrm{m}},
\end{equation}
where $Y_{\text{form}}=10^{-4}\,M_{\odot}^{-1}\,\mathrm{Gyr}^{-1}$ is the compact-binary formation efficiency per unit cluster mass, $M_{\mathrm{cl}} = 3.2 \times 10^7 M_{\odot}$ and $n_{\mathrm{cl}}=0.02\,\mathrm{Mpc}^{-3}$ is the number density of nuclear star clusters. $f_{\mathrm{m}}$ is BHBs merger fraction from our numerical simulations. Substituting these values yields a merger rate of $\mathcal{R} \simeq 16\,\mathrm{Gpc}^{-3}\,\mathrm{yr}^{-1}$.

The merger rate of BHBs in galactic nuclei has been investigated through several distinct dynamical channels, which can be broadly classified into three categories: (1) secular evolution driven by coherent dynamical processes, including SMBH-induced ZLK oscillations and perturbations from the smooth nuclear star cluster potential; (2) non-secular evolution driven by discrete dynamical encounters, such as stochastic stellar flybys; and (3) gas-assisted channels associated with AGN disks.

Previous studies focusing primarily on secular processes have generally predicted relatively modest BHB merger rates in galactic nuclei. For example, the SMBH-induced ZLK channel alone yields merger fractions of $\sim7\%$ and corresponding merger rates of $\sim1$--$3\mathrm{Gpc}^{-3}\mathrm{yr}^{-1}$ (e.g., \citealt{Hoang.2018ApJ...856..140H}). The inclusion of a smooth stellar cluster potential can further modify the secular evolution by introducing additional apsidal precession. However, models that neglect stochastic stellar encounters typically predict lower merger efficiencies, with rates of only $\sim0.1$--$0.2\mathrm{Gpc}^{-3}\mathrm{yr}^{-1}$ (e.g., \citealt{Hamilton.2019ApJ...881L..13H}). Other studies incorporating SMBH perturbations while neglecting additional environmental effects have reported rates spanning $0.002$--$0.6\mathrm{Gpc}^{-3}\mathrm{yr}^{-1}$ (e.g., \citealt{2012ApJ...757...27A}) and $0.3$--$5\mathrm{Gpc}^{-3}\mathrm{yr}^{-1}$ (e.g., \citealt{Wang.2021ApJ...917...76W}). Conversely, departures from spherical symmetry in nuclear star clusters can enhance secular eccentricity excitation and lead to higher merger rates, reaching $0.6$--$15\mathrm{Gpc}^{-3}\mathrm{yr}^{-1}$ in the presence of an SMBH (e.g., \citealt{2017ApJ...846..146P}).

For comparison, gas-assisted environments represent another efficient pathway for BHB mergers. Models involving BHB migration and interaction in AGN disks predict merger rates of $\sim0.02$--$100\mathrm{Gpc}^{-3}\mathrm{yr}^{-1}$, although these estimates remain highly uncertain due to the complex interplay between gas dynamics, disk structure, and AGN population properties (e.g., \citealt{Tagawa.2020ApJ...898...25T,2018ApJ...866...66M,2017ApJ...835..165B,2017MNRAS.464..946S,2019ApJ...876..122Y}). A broader discussion of BHB merger rates from dynamical channels in galactic nuclei can be found in the review (e.g., \citealt{Mandel.2022LRR....25....1M}).

Our results demonstrate that the combined action of secular dynamics and stellar flybys significantly enhances the merger efficiency compared with either mechanism operating independently. In particular, stellar flybys enable BHBs in the intermediate regions of galactic nuclei, where purely ZLK-driven mergers are inefficient to reach the merger regime within a Hubble time. This stochastic enhancement increases the overall merger fraction and leads to a predicted rate of $\mathcal{R}\approx16\mathrm{Gpc}^{-3}\mathrm{yr}^{-1}$, exceeding most previous estimates based solely on secular evolution.

\section{Conclusion and Discussion}\label{sec:IV}
\subsection{Conclusion}
In this study, we have investigated the long term evolution of BHBs in galactic nuclei by incorporating several secular and non-secular effects, including the tidal influence of the central SMBH, the potential of the nuclear star cluster, and SMBH induced relativistic precession, together with the stochastic perturbations from stellar flybys. By evolving a large population of hierarchical triples over a Hubble time, we have quantified the relative efficiency of these competing processes in driving BHBs toward either merger or disruption. Our main findings are summarized as follows.
\begin{itemize}

    \item We find that the BHB evolution in galactic nuclei is governed by two dominant dynamical drivers: the secular ZLK oscillations from the SMBH, and the nonsecular perturbations from stellar flybys. The pure ZLK scenario predominantly operates in the immediate vicinity of the SMBH. The stellar flybys opens an additional merger channel at intermediate galactocentric distances ($0.02$–$0.1\,\mathrm{pc}$).

    \item Regardless of the binary's galactocentric distance, the long term evolution of BHBs is always influenced by stellar flybys to varying degrees. These cumulative perturbations drive binaries to extreme eccentricities, triggering both mergers and disruptions. The environmental impact is substantial: among initially stable systems, $\sim 24.8\%$ merge and $\sim 55.6\%$ are disrupted, leaving only a small surviving over the Hubble time. Note that this channel increases the merger fraction by a factor of \(\sim3\) compared with the flyby-free case.

    \item We infer a merger rate density of $\mathcal{R} \simeq 16\,\mathrm{Gpc}^{-3}\,\mathrm{yr}^{-1}$, which significantly exceeds previous theoretical estimates. This suggests that galactic nuclei, when fully accounting for environmental effects, may constitute a major—and possibly dominant—channel for BHB mergers.

    \item Most merging systems enter the LIGO band with eccentricities $e \lesssim 10^{-3}$, yet retain detectable eccentricities at lower frequencies ($\sim 5\,\mathrm{Hz}$), offering a clear observational target for next‑generation detectors.

\end{itemize}

\subsection{Discussion}
Our framework captures the essential dynamical processes governing BHB evolution in galactic nuclei, including secular interactions, relativistic effects, and stochastic stellar perturbations. Nevertheless, several aspects of the physical modeling remain simplified and motivate further investigation.

First, the environmental properties of galactic nuclei are treated in an idealized manner. We adopt a spherically symmetric and time-independent nuclear star cluster potential, whereas realistic nuclei can exhibit non-spherical structures, mass segregation, and secular evolution. These features can modify the apsidal and nodal precession of BHB orbits and thereby regulate the efficiency of eccentricity excitation (e.g., \citealt{2013degn.book.....M}). In particular, triaxial nuclear potentials can sustain centrophilic stellar orbits and introduce additional secular torques, potentially altering the dynamical evolution of binaries orbiting SMBHs (e.g., \citealt{2017ApJ...846..146P}). 
Furthermore, the nuclear star cluster (NSC) may have been less massive and less dense at earlier epochs, before relaxation and mass segregation substantially reshaped its inner regions. A lower local stellar density would imply less frequent flyby encounters. According to Eq. (\ref{eq:10}), the flyby timescale scales inversely with stellar density at fixed encounter cross section and velocity. Thus, if the NSC had half its present-day mass but the same density-profile shape and scale radius, the flyby timescale at a given radius would be approximately twice as long, provided that the encounter velocity remained similar. This approximation is most applicable in the SMBH-dominated region. Such evolution could alter the cumulative effect of stellar encounters, particularly in the inner NSC where encounters are frequent. However, the merger fraction need not scale linearly with density: flybys can both induce mergers and disrupt binaries, while SMBH-driven secular torques remain active even when encounters are less frequent. The net effect on the merger rate therefore requires a time-dependent model of the stellar density, velocity dispersion, and compact-object population. We leave this treatment to future work. Combined with the evolution of the BHB formation rate and NSC population, and accounting for merger delays, such a model would allow predictions of the BHB merger rate as a function of redshift.
Moreover, we assume a fixed magnitude and orientation of the SMBH spin, while realistic SMBH spin distributions are expected to depend on their accretion histories and merger-driven growth (e.g., \citealt{Liu:2019tqr}). A more comprehensive exploration of nuclear structures and self-consistent SMBH spin evolution will be considered in future studies.

Second, several aspects of the stochastic perturbation and BHB formation processes are not included in our current population synthesis model. Stellar flybys are modeled using an approximate encounter cross section and an idealized stellar density distribution. In realistic nuclear environments, correlated encounters, massive perturbers, and resonant relaxation can modify the diffusion of binary angular momentum and eccentricity (e.g., \citealt{Rauch:1996fb,Hopman:2006qr,Perets:2007nc}), potentially enhancing or suppressing the efficiency of eccentricity excitation. Furthermore, we adopt simplified initial conditions and focus on the evolution of first-generation (1G) BHBs, without explicitly modeling their formation channels. In realistic galactic nuclei, BHBs may originate from isolated stellar evolution, dynamical assembly through single--single and binary--single interactions, or previous generations of hierarchical mergers (e.g., \citealt{Rodriguez:2015oxa,Antonini:2016gqe,Mapelli:2016vca}). Repeated mergers can further produce second-generation black holes with distinctive mass and spin distributions (e.g., \citealt{Gerosa:2017kvu}). Incorporating self-consistent BHB formation pathways, exchange interactions, and multiple generations of mergers will be an important direction for future studies (e.g., \citealt{Liu:2025yok}).


Third, our model assumes a gas-free NSC and does not include the effects of intermittent AGN activity (e.g., \citealt{2015MNRAS.451.2517S,1982MNRAS.200..115S}). During active phases, interactions with an accretion disk may accelerate BHB mergers and deplete part of the binary population available for subsequent mergers through the quiescent ZLK+flyby channel (e.g., \citealt{2022MNRAS.517.5827F,2019ApJ...876..122Y}). A self-consistent treatment of these effects requires modeling the properties and evolution of the disk, the duration and recurrence of AGN episodes, and the depletion and replenishment of the BHB population, which is beyond the scope of this work. As a comparison, we examine the merger statistics over different evolution times in our gas-free model. The cumulative merger fractions within 10 Myr, 100 Myr, 1 Gyr, and 10 Gyr are approximately 0.05, 0.12, 0.20, and 0.25, respectively. Under the formation-rate prescription adopted in Eq. (\ref{eq:20}), these fractions correspond to merger-rate estimates of approximately 3.2, 7.7, 12.8, and 16 Gpc$^{-3}$ yr$^{-1}$. These results illustrate how the merger contribution builds up over different timescales in a gas-free NSC and provide a reference for future studies incorporating intermittent AGN activity.

Overall, our study underscores the critical need to include both secular dynamics and non-secular environmental perturbations when modeling BHB populations in galactic nuclei. Future work should incorporate more realistic cluster models and self-consistent population synthesis. Meanwhile, next-generation GW observations, particularly at low frequencies, will provide critical tests of these dynamical scenarios.

\begin{acknowledgments}
B.L. acknowledges support from the National Natural Science Foundation of China (grant No. 12673087 and 12433008) and National Key Research and Development Program of China (No. 2023YFB3002502).
In this paper, we utilized the high-performance computing cluster
"Zimo" from the Purple Mountain Observatory of the Chinese Academy of Sciences, which is equipped with GPU cards made in China. J.S. is supported by
the Villum Fonden grant No. 29466, and by the ERC Starting Grant no.
101043143 – BlackHoleMergs. The Center of Gravity is a Center of Excellence
funded by the Danish National Research Foundation under grant No. 184.  
\end{acknowledgments}

\appendix \label{App:A}
\renewcommand{\thesection}{\Alph{section}}

\section{Evolution equations}\label{app:A}

We adopt the formalism and notation of \citet{Liu.2015MNRAS.447..747L}, and describe the orbital geometry using an orthonormal triad $\{\hat{\boldsymbol{e}}, \hat{\boldsymbol{v}}, \hat{\boldsymbol{j}}\}$. Here, $\hat{\boldsymbol{e}}$ points toward the pericenter, $\hat{\boldsymbol{v}}$ lies in the orbital plane perpendicular to $\hat{\boldsymbol{u}}$, and $\hat{\boldsymbol{j}}=\hat{\boldsymbol{e}} \times \hat{\boldsymbol{v}}$ is the unit vector along the orbital angular momentum. This triad is used to represent both the inner and outer orbits.

Within this framework, the dimensionless angular momentum and eccentricity vectors are defined as
\begin{equation}
\boldsymbol{j}=\sqrt{1-e^2}\,\hat{\boldsymbol{j}}, \quad \boldsymbol{e}=e\,\hat{\boldsymbol{e}},
\end{equation}
which provide a convenient vectorial description of the secular dynamics.

The double-averaged Hamiltonian at quadrupole order can be written as (e.g., \citealt{Hamilton.2019MNRAS.488.5489H,Hamilton.2019MNRAS.488.5512H})
\begin{equation}
{\overline{\left\langle H_1\right\rangle}}_M=C H_1^*,
\end{equation}
where $C=A a_{\text{in}}^2 / 8$ and
\begin{equation}\label{eq:A3}
H_1^*=2\left[1-\frac{3}{2} e_{\mathrm{in}}^2(5 \Gamma-1)-3 \Gamma\left(\boldsymbol{j}_{\mathrm{in}} \cdot \boldsymbol{j}_{\mathrm{out}}\right)^2+15 \Gamma\left(\boldsymbol{e}_{\mathrm{in}} \cdot \boldsymbol{j}_{\mathrm{out}}\right)^2\right].
\end{equation}
Here, the coefficients $A$ and $\Gamma$ encode the influence of the external tidal field and are determined by the torus-averaged cluster potential via
\begin{equation} \label{eq:A4}
A \equiv \bar{\Phi}_{z z}+\bar{\Phi}_{x x}, \quad B \equiv \bar{\Phi}_{z z}-\bar{\Phi}_{x x}, \quad \Gamma \equiv \frac{B}{3A}.
\end{equation}

Assuming an underlying Hernquist profile for the stellar cluster, the orbit-averaged potential yields
\begin{equation}\label{eq:A5}
\Gamma=\frac{M_{\mathrm{cl}} r^2(b_{\mathrm{Hern}}+3 r)+3 m_3(b_{\mathrm{Hern}}+r)^3}{3\left[M_{\mathrm{cl}} r^2(3 b_{\mathrm{Hern}}+r)+m_3(b_{\mathrm{Hern}}+r)^3\right]},
\end{equation}
\begin{equation}\label{eq:A6}
A=\frac{G M_{\mathrm{cl}}(3 b_{\mathrm{Hern}}+r)}{2 r(b_{\mathrm{Hern}}+r)^3}+\frac{G m_3}{2 r^3}.
\end{equation}
The parameter $\Gamma$ governs the phase-space structure of the inner binary’s orbital dynamics, whereas $A$ serves as an overall scaling factor that sets the timescale for secular eccentricity oscillations (e.g., \citealt{Hamilton.2019MNRAS.488.5489H,Hamilton.2019MNRAS.488.5512H}). 

The corresponding secular timescale is
\begin{equation}\label{eq:A7}
T_{\text{Hern+SMBH}}=\frac{2 \sqrt{G (m_1+m_2)}}{3 A a_{\mathrm{in}}^{3 / 2}},
\end{equation}
which characterizes the rate of angular momentum exchange induced by the combined SMBH--cluster potential. Notably, $\Gamma$ quantifies deviations from spherical symmetry, and thus controls the strength and morphology of Lidov–Kozai–like oscillations in this generalized potential. 

From the Hamiltonian (Eq. (\ref{eq:A3})), the secular equations of motion in vector form are obtained as
\begin{equation}\label{eq:A8}
\begin{aligned}
\frac{\mathrm{d} \boldsymbol{L}_{\mathrm{in}}}{\mathrm{~d} t}=&\frac{3L_{\text {in }}}{4 T_{\text{Hern+SMBH}}\sqrt{1-e_{\mathrm{in}}^2}}\left[\Gamma \left(1-e_{\mathrm{in}}^2\right)\left(\hat{\boldsymbol{j}}_{\mathrm{in}} \cdot \hat{\boldsymbol{j}}_{\mathrm{out}}\right) \hat{\boldsymbol{j}}_{\mathrm{in}} \times \hat{\boldsymbol{j}}_{\mathrm{out}}-5 \Gamma\left(\boldsymbol{e}_{\mathrm{in}} \cdot \hat{\boldsymbol{j}}_{\mathrm{out}}\right) \boldsymbol{e}_{\mathrm{in}} \times \hat{\boldsymbol{j}}_{\mathrm{out}}\right]\\
\frac{\mathrm{d} \boldsymbol{e}_{\mathrm{in}}}{\mathrm{~d} t}=&\frac{3\sqrt{1-e_{\mathrm{in}}^2}}{4 T_{\text{Hern+SMBH}}}\left[\Gamma\left(\hat{\boldsymbol{j}}_{\mathrm{in}} \cdot \hat{\boldsymbol{j}}_{\mathrm{out}}\right) \boldsymbol{e}_{\mathrm{in}} \times \hat{\boldsymbol{j}}_{\mathrm{out}}+\frac{1}{2}(5 \Gamma-1) \hat{\boldsymbol{j}}_{\mathrm{in}} \times \boldsymbol{e}_{\mathrm{in}}-5 \Gamma\left(\boldsymbol{e}_{\mathrm{in}} \cdot \hat{\boldsymbol{j}}_{\mathrm{out}}\right) \hat{\boldsymbol{j}}_{\mathrm{in}} \times \hat{\boldsymbol{j}}_{\mathrm{out}}\right]\\
\frac{\mathrm{d} \boldsymbol{L}_{\mathrm{out}}}{\mathrm{~d} t}=&\frac{3L_{\mathrm{in}}}{4 T_{\text{Hern+SMBH}} \sqrt{1-e_{\mathrm{in}}^2}} \left[\Gamma\left(1-e_{\mathrm{in}}^2\right)\left(\hat{\boldsymbol{j}}_{\mathrm{in}} \cdot \hat{\boldsymbol{j}}_{\mathrm{out}}\right) \hat{\boldsymbol{j}}_{\mathrm{out}} \times \hat{\boldsymbol{j}}_{\mathrm{in}}-5 \Gamma\left(\boldsymbol{e}_{\mathrm{in}} \cdot \hat{\boldsymbol{j}}_{\mathrm{out}}\right) \hat{\boldsymbol{j}}_{\mathrm{out}} \times \boldsymbol{e}_{\mathrm{in}}\right]\\
\frac{\mathrm{d} \boldsymbol{e}_{\mathrm{out}}}{\mathrm{~d} t}=&\frac{3}{4 T_{\text{Hern+SMBH}} \sqrt{1-e_{\mathrm{out}}^2}\sqrt{1-e_{\mathrm{in}}^2}} \frac{L_\text{in}}{L_\text{out}}\left[\Gamma\left(1-e_1^2\right)\left(\hat{\boldsymbol{j}}_{\mathrm{in}} \cdot \hat{\boldsymbol{j}}_{\mathrm{out}}\right) \boldsymbol{e}_{\mathrm{out}} \times \hat{\boldsymbol{j}}_{\mathrm{in}}-5 \Gamma\left(\boldsymbol{e}_{\mathrm{in}} \cdot \hat{\boldsymbol{j}}_{\mathrm{out}}\right) \boldsymbol{e}_{\mathrm{out}} \times \boldsymbol{e}_{\mathrm{in}}\right.\\ 
&\left.-\left[\frac{1}{2}-\frac{3}{4}(5 \Gamma-1) e_{\mathrm{in}}^2+\frac{25}{2} \Gamma\left(\boldsymbol{e}_{\mathrm{in}} \cdot \hat{\boldsymbol{j}}_{\mathrm{out}}\right)^2-\frac{5}{2} \Gamma\left(1-e_{\mathrm{in}}^2\right)\left(\hat{\boldsymbol{j}}_{\mathrm{in}} \cdot \hat{\boldsymbol{j}}_{\mathrm{out}}\right)^2\right] \hat{\boldsymbol{j}}_{\mathrm{out}} \times \boldsymbol{e}_{\mathrm{out}}\right].
\end{aligned}
\end{equation}

Here, $\boldsymbol{L}_{\mathrm{in}} = L_{\mathrm{in}} \hat{\boldsymbol{L}}_{\mathrm{in}}$ ($\hat{\boldsymbol{L}}_{\mathrm{in}}=\hat{\boldsymbol{j}}_{\mathrm{in}}$) and $\boldsymbol{L}_{\mathrm{out}} = L_{\mathrm{out}} \hat{\boldsymbol{L}}_{\mathrm{out}}$ ($\hat{\boldsymbol{L}}_{\mathrm{out}}=\hat{\boldsymbol{j}}_{\mathrm{out}}$) denote the angular momenta vectors of the inner and outer orbits, respectively. Their magnitudes are given by:
\begin{equation}
\begin{aligned}
& L_\text{in}=\mu_\text{in} \sqrt{G\left(m_1+m_2\right) a_\text{in}}, \\
& L_\text{out}=\mu_\text{out} \sqrt{G\left(m_1+m_2+m_3\right) a_\text{out}},
\end{aligned}
\end{equation}
with reduced masses
\begin{equation}
\mu_\text{in}=\frac{m_1 m_2}{m_1+m_2}.
\end{equation}

\paragraph{Relativistic corrections.}
General relativistic (GR) effects introduce additional precession and dissipation terms. The leading-order (1PN) pericenter precession of the inner binary is
\begin{equation}\label{eq:A11}
\left.\frac{d \boldsymbol{e}_{\mathrm{in}}}{d t}\right|_{\mathrm{GR}}=\dot{\omega}_{\mathrm{GR}} \hat{\boldsymbol{L}}_{\mathrm{in}} \times \boldsymbol{e}_{\mathrm{in}},
\end{equation}
where the precession rate is
\begin{equation}\label{eq:A12}
\dot{\omega}_{\mathrm{GR}}=\frac
{3 G^{3/2}(m_1+m_2)^{3/2}}{c^2 a_{\mathrm{in}}^{5/2}\left(1-e_{\mathrm{in}}^2\right)}.
\end{equation}

At the dissipative level, we incorporate GW emission at 2.5PN order following Peters (e.g., \citealt{Peters:1963ux,Peters:1964zz}):
\begin{equation} \label{eq:A13}
\frac{\mathrm{d} \boldsymbol{L}_{\mathrm{in}}}{\mathrm{~d} t}=- \frac{32}{5} \frac{G^{7 / 2}}{c^5} \frac{\mu^2}{a_{\mathrm{in}}^{7 / 2}}\left(m_1+m_2\right)^{5 / 2} \frac{1+\frac{7}{8} e_{\mathrm{in}}^2}{\left(1-e_{\mathrm{in}}^2\right)^2}\hat{\boldsymbol{L}}_{\mathrm{in}},
\end{equation}
\begin{equation}\label{eq:A14}
\frac{\mathrm{d} \boldsymbol{e}_{\mathrm{in}}}{\mathrm{~d} t}=- \frac{304 G^3 m_1 m_2\left(m_1+m_2\right)}{15 c^5 a_{\mathrm{in}}^4\left(1-e_{\mathrm{in}}^2\right)^{5 / 2}}\left(1+\frac{121}{304} e_{\mathrm{in}}^2\right)\hat{\boldsymbol{e}}_{\mathrm{in}}.
\end{equation}

\section{GR effects of SMBH}

We further include relativistic corrections associated with the central SMBH spin and the outer orbit, which can qualitatively modify the secular dynamics.

(i) \textit{Effect I: Lense–Thirring precession of the outer orbit.}
\begin{equation}
\begin{aligned}
& \left.\frac{d \boldsymbol{L}_{\text {out }}}{d t}\right|_{\mathrm{L}_{\text {out }} \mathrm{S}_3}=\Omega_{\mathrm{L}_{\text {out }} \mathrm{S}_3} \hat{\boldsymbol{S}}_3 \times \boldsymbol{L}_{\text {out }}, \\
& \left.\frac{d \boldsymbol{e}_{\text {out }}}{d t}\right|_{\mathrm{L}_{\text {out }} \mathrm{S}_3}=\Omega_{\mathrm{L}_{\text {out }} \mathrm{S}_3} \hat{\boldsymbol{S}}_3 \times \boldsymbol{e}_{\text {out }} 
 -3 \Omega_{\mathrm{L}_{\text {out }} \mathrm{S}_3}\left(\hat{\boldsymbol{L}}_{\text {out }} \cdot \hat{\boldsymbol{S}}_3\right) \hat{\boldsymbol{L}}_{\text {out }} \times \boldsymbol{e}_{\text {out }}.
\end{aligned}
\end{equation}
The orbit-averaged precession rate is
\begin{equation} 
\Omega_{\mathrm{L}_{\mathrm{out}} \mathrm{S}_3}=\frac{G S_3\left(4+3 (m_1+m_2) / m_3\right)}{2 c^2 a_{\mathrm{out}}^3\left(1-e_{\mathrm{out}}^2\right)^{3 / 2}}.
\end{equation}

(ii) \textit{Effect II: de Sitter–like precession of the inner orbit.}
\begin{equation}
\begin{aligned}
&\left.\frac{d \boldsymbol{L}_{\text {in }}}{d t}\right|_{\mathrm{L}_{\text {in }} \mathrm{L}_{\text {out }}}=\Omega_{\mathrm{L}_{\text {in }} \mathrm{L}_{\text {out }}} \hat{\boldsymbol{L}}_{\text {out }} \times \boldsymbol{L}_{\text {in }},\\
&\left.\frac{d \boldsymbol{e}_{\text {in }}}{d t}\right|_{\mathrm{L}_{\text {in }} \mathrm{L}_{\text {out }}}=\Omega_{\mathrm{L}_{\text {in }} \mathrm{L}_{\text {out }}}\hat{\boldsymbol{L}}_{\text {out }} \times \boldsymbol{e}_{\text {in }},
\end{aligned}
\end{equation}
with
\begin{equation}
\Omega_{\mathrm{L}_{\mathrm{in}}\mathrm{L}_{\mathrm{out}}}=\frac{3}{2} \frac{G\left(m_3+\mu_{\mathrm{out}} / 3\right) \left(G M_{\mathrm{tot}} / a_{\mathrm{out}}^3\right)^{1 / 2}}{c^2 a_{\mathrm{out}}\left(1-e_{\mathrm{out}}^2\right)}.
\end{equation}

(iii) \textit{Effect III: Lense–Thirring precession of the inner orbit.}
\begin{equation}
\begin{aligned}
& \left.\frac{d \boldsymbol{L}_{\text {in }}}{d t}\right|_{\mathrm{L}_{\text {in }} \mathrm{S}_3}=\Omega_{\mathrm{L}_{\text {in }} \mathrm{S}_3} \hat{\boldsymbol{S}}_3 \times \boldsymbol{L}_{\text {in }}, \\
& \left.\frac{d \boldsymbol{e}_{\text {in }}}{d t}\right|_{\mathrm{L}_{\text {in }} \mathrm{S}_3}=\Omega_{\mathrm{L}_{\text {in }} \mathrm{S}_3} \hat{\boldsymbol{S}}_3 \times \boldsymbol{e}_{\text {in }},
\end{aligned}
\end{equation}
with
\begin{equation}
\Omega_{\mathrm{L}_{\mathrm{in}} \mathrm{S}_3}=\frac{G S_3}{2 c^2 a_{\mathrm{out}}^3\left(1-e_{\mathrm{out}}^2\right)^{3 / 2}}. 
\end{equation}

\bibliography{sample701}{}
\bibliographystyle{aasjournal}

\end{document}